# Reconstruction of the interatomic forces from dynamic Scanning Transmission Electron Microscopy data

M. Chakraborty,[1] M. Ziatdinov,[2] O. Dyck,[2] S. Jesse,[2] A. D. White,[1,a] and Sergei V. Kalinin[2,b]

[1] Department of Chemical Engineering, University of Rochester, Rochester, NY 14627, USA

[2] The Center for Nanophase Materials Sciences, Oak Ridge National Laboratory, Oak Ridge, TN 37831, USA

We explore the possibility for reconstruction of the generative physical models describing interactions between atomic units in solids from observational electron microscopy data. Here, scanning transmission electron microscopy (STEM) is used to observe the dynamic motion of Si atoms at the edge of monolayer graphene under continuous electron beam illumination. The resulting time-lapsed STEM images represent the snapshots of observed chemical states of the system. We use two approaches: potential of mean force (PMF) calculation using a radial distribution function (RDF) and a direct fitting of the graphene-Si interatomic pair-wise potentials with force matching, to reconstruct the force fields in the materials. These studies lay the foundation for quantitative analysis of materials energetics from STEM data through the sampling of the metastable states in the chemical space of the system.

[a] andrew.white@rochester.edu
[b] sergei2@ornl.gov

**Introduction**

Introduction of aberration correction in Scanning Transmission Electron Microscopy (STEM) has propelled this method to a technique of choice for characterizing a broad range of materials including metals and semiconductors, oxides, 2D materials, and many others.[1-4] Beyond multiple qualitative studies visualizing structure of interfaces and localized and extended defects, recent improvements in spatial resolution and the stability of STEM instrumentation have enabled quantitative measurements of atom position with picometer precision,[5] with further advances enabled by segmented detectors and 4D STEM such as sub-picometer precision strain mapping,[6] sub-angstrom charge-density mapping,[7] electron ptychography for improved spatial resolution,[8] and differential phase-contrast imaging.[9] This quantitative structural STEM imaging now allows insight into the materials structure and properties that previously was achievable only on average via diffraction-based methods, with the added advantage of element-, site-, and even charge/valence sensitivity.[10]

This rapid growth in quantitative STEM has opened fundamentally new opportunities for exploring physics and chemistry of materials based on local structural measurements. Jia (via TEM)[11-13] and Chisholm (via STEM)[14] following the earlier work of Pan[15] demonstrated direct probing of the symmetry breaking distortions in perovskites, enabling direct mapping of the polarization order parameter fields. This approach was further extended by Jia[16] and He[17] to probe octahedra tilts in the image plane and, via column shape analysis,[18] in the beam directions. These studies have provided insight into ferroelectricity and screening phenomena at the oxide interfaces, emergence of topological defects in ferroelectric superlattices, etc. Mapping of the order parameter fields have now become de facto standard in the field.[19-24]

Mapping of the mesoscopic order parameter fields can further be extended to extract the specific aspects of materials physics via matching with the Ginzburg-Landau type models. In one such approach, the analytic solutions for well-defined geometries with unknown boundary and gradient terms can be fitted to the atomically resolved fields to yield the value of the relevant materials constants.[25,26] Alternatively, the gradient coupling terms such as flexoelectric constant can be determined from theory-experiment matching.[27]

However, mesoscopic order parameter fields can be defined only for the systems with continuous lattices and small number of point and extended defects. Correspondingly, the potential of high-resolution STEM data to provide insight into the physics and chemistry of the systems with significant chemical disorder remain largely unexplored. Recently, it was proposed that such data can be mined to define a correlative picture of material structure, for example libraries of structural units and defects.[28,29] This information can be further used to build the generative physical models. For lattice models, Vlcek et al.[30-33] has demonstrated an approach to extract the interaction Hamiltonians from the images of chemically-disordered systems using a statistical distance minimization method. However, until now there have been no attempts to analyze the generative models behind the chemical transformations in STEM associated with the change of chemical bonding patterns.

Here, we demonstrate the analysis of dynamic STEM data to extract interaction potentials from the observations of atomic units. We use dynamic STEM data to acquire multiple snapshots of the dynamic process of chemical evolution in the Si-graphene system and use these to reconstruct the possible interaction potentials.

**Results and discussion**

The graphene samples were fabricated using a wet transfer from the Cu growth substrate to a STEM sample grid and an oxygen baking procedure for cleaning, as detailed elsewhere.[34] The dynamic observations were performed in a Nion UltraSTEM US200 operating at 100 kV accelerating voltage. The beam current was set to 10 pA and images were acquired with the high-angle annular dark field (HAADF) detector to preserve Z-contrast.[35-37] In this imaging mode, heavier atoms appear brighter allowing for a straight-forward interpretation of image contrast and identification of all atoms. An area of the graphene sample was found where some contaminant material, comprised of mostly amorphous C and Si atoms, was next to an atomically pristine portion of the graphene lattice. An image sequence was then recorded over ~56 mins, capturing the sample evolution under the influence of the 100 kV e-beam in a 200 frame data stack (64 μs pixel dwell time, 31 pm pixel size, 16 nm field of view, 17 s per frame). Figure 1 shows a selection of the acquired frames with the total accumulated dose indicated in the upper right (calculated as beam current * frame time * # of frames / frame area). We note that the atomic processes involved in the observed evolution occur much faster than can be directly captured in each frame. We observe a significant doping of the graphene with the bright Si atoms at the edge of the contaminated region, consistent with previous observations of e-beam driven doping.[38] This doping and possible chemical interaction with other species in the contamination (e.g. O) weakens the bonding of the C atoms and promotes structural degradation in this region. We observe a complete separation open between the contaminated region and the graphene lattice. As atoms are continually sputtered from the lattice we also observe the appearance and movement of defects as well as the retraction of the graphene edge as the total number of C atoms decreases.

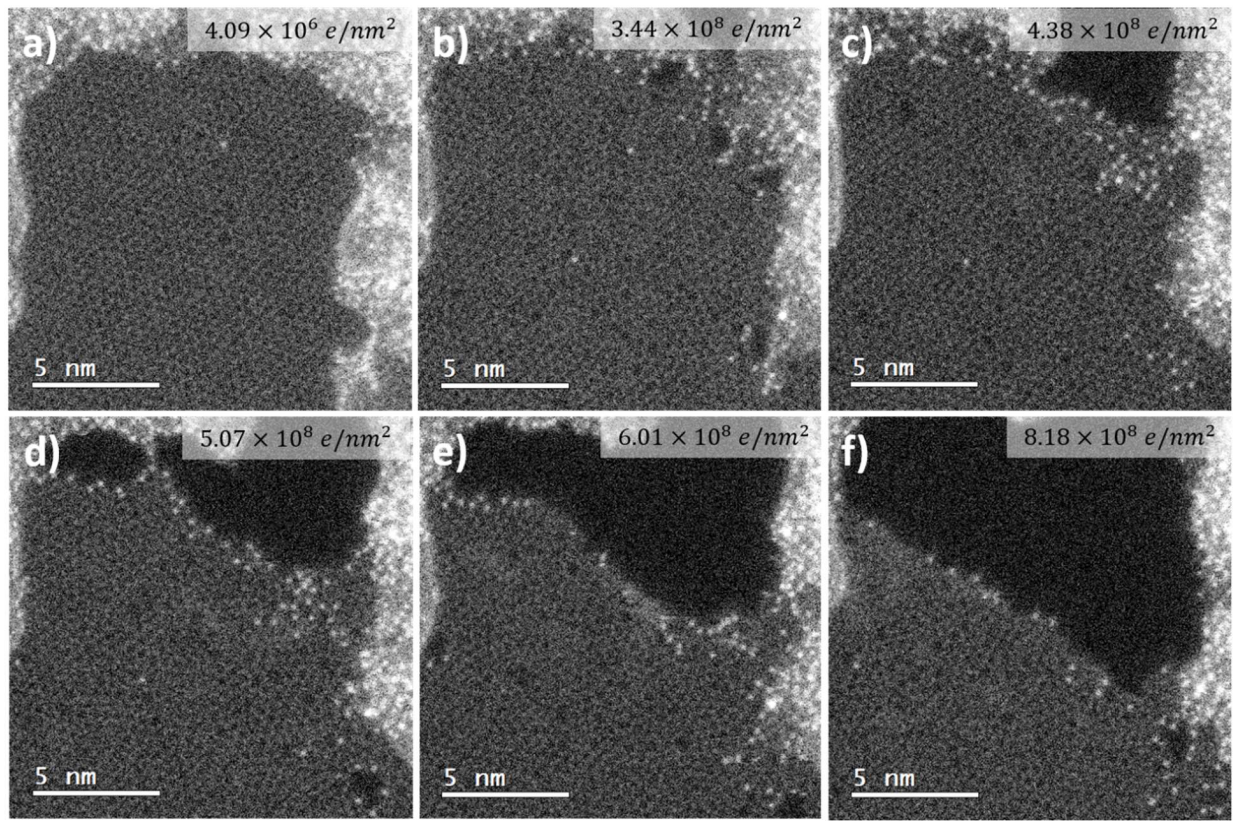

**Figure 1 HAADF-STEM time series of graphene degrading under electron irradiation at 100 kV.** a) Initial state where the mostly pristine graphene lattice can be seen to be surrounded by an area of contamination consisting of primarily amorphous C and Si atoms. b)-f) structural evolution of the graphene lattice during scanning of the electron beam at time steps 1428 s (b), 1819 s (c), 2108 s (d), 2499 s (e), 3400 s (f). The accumulated total electron dose at each point in time is indicated in the upper right corner of each image.

The e-beam directed atomic evolution process is non-uniform in time. The graphene lattice is more stable at the beginning (i.e. pristine graphene is much more robust) and transformation from (a) to (b) proceeds over 84 frames (out of 200). The lattice degradation is apparently accelerated at the interface with the contamination, which could be due to chemical reactions with contaminant atoms driven by the e-beam. Finally, the most interesting phenomena are observed in the bulk of the graphene region. During continuous electron beam exposure, the graphene does not nucleate multi-atomic voids (holes). Rather a continuous generation of point defects such as Stone-Wales defects (5-7 rings), etc. is observed. Once formed, the defects apparently migrate to the edge, while preserving the integrity of the bulk. The edge recession observed from (c) to (f) is therefore driven by both defect migration and edge sputtering. This in turn suggests that the majority of the atomic dynamics observed during STEM imaging corresponds to the reversible transitions between possible defect states in the material, and the process is almost ergodic. In other words, the electron beam activates the Si-graphene system allowing it to relax into possible metastable states, and in such a way effectively samples the allowed chemical states of the system which can be energetically reached through the impartation of the beam energy.

We further note that both the transit time of the electron through the graphene sheet (~attoseconds) and the lifetimes of the electronic excitations in graphene are well below the time scale of individual images. Therefore, the excitations that produce chemical changes in the system are essentially a delta function in time, and observed dynamics corresponds to the chemical relaxation processes. Therefore, the observed atomic configurations correspond to the snapshots of the possible chemical states of the graphene-Si system.

To identify positions of all atoms in every frame of the experimental movie we utilized a deep fully convolutional neural network (FCNN). The application of FCNN to atom-resolved STEM data is described elsewhere.[39] Briefly, the network was trained using simulated data to remove noise from raw experimental images and to separate image pixels associated with different atomic species into different classes. The structures used for model training were obtained from first-principles calculations of the relaxed geometries from the extended library of atomic defects in graphene under different strain.[28] The relaxed coordinates were then used as an input into MultiSlice algorithm[40] to produce the simulated STEM images, which were corrupted by different types and levels of noise. The model training was done by utilizing Google's Tensor Processing Units. The FCNN output is a set of well-defined circular features on a uniform background (see SI). To convert the FCNN output into atomic coordinates we used a simple center of mass measurement on the extracted features.

These data hence provide the information on possible configurations in the Si-graphene system. Due to the extremely rapid character of electron beam induced processes and associated energy relaxation in graphene and low yield of electron beam introduced changes, these images represent multiple realizations of the metastable atomic configurations in the chemical space in the system. Each of these in turn corresponds to the zero force acting on individual atoms. Here, we explore whether this information can be used to reconstruct the force fields acting between atoms.

As a context to these studies, there have been previous computational studies aimed at studying interactions in graphene. A study by Inui and Iwasaki investigated the interaction energy between a graphene sheet and a silicon substrate.[41] The pairwise nonbonded interactions, modeled as Lennard-Jones (LJ) potential, were $\sigma = 3.629$ Å and $\epsilon = 8.91\ meV$ for the C-Si and the C-C parameters are given Table 1. In a LJ potential, $\sigma$ refers to the distance at which the pairwise potential is 0 and $\epsilon$ refers to the depth of the potential well indicating the maximum strength of the attractive interaction. We also include the list of LJ parameters for non-bonded carbon atom interactions used for graphene modeling as reported in a review.[42] The Lennard-Jones is perhaps the crudest assumption possible, especially since it cannot reproduce the lattice structure of graphene, but it can provide a starting point for comparison.

**Table 1.** Lennard-Jones parameters for pairwise interactions between carbon atoms in graphene sheet. Some parameters taken from Pykal et al.[42]

| Reference | $\sigma$ (Å) | $\epsilon$ (meV) |
|---|---|---|
| Inui and Iwasaki.[41] | 3.431 | 4.55 |
| Parm 99[43] | 3.39967 | 3.7 |
| OPLS[44] | 3.5500 | 3.0 |
| CHARMM27[45] | 3.5505 | 3.0 |
| Ulbricht et al.[46] | 3.78108 | 2.63 |
| Girifalco et al.[47] | 3.41214 | 2.39 |
| Cheng and Steele[48] | 3.39967 | 2.42 |
| COMPASS[49] | 3.48787 | 2.9 |

Besides non-bonded interaction energies, different reactive energies in graphene, like bond-breaking, vacancy formation, diffusion, and merging and deformation energies, have been previously reported. Sun et al.[50] studied C 1s binding energy of carbon in different materials using DFT and reported C 1s binding energy to be 284.80 eV for interior atoms of graphene. Machine learning[51], *ab initio* calculations[51] and quantum based Morse potentials[52] have also been used to estimate the aforementioned reactive energies of graphene carbon. Based on previous studies of graphene, dissociation energy was estimated to be 8.34 eV[52], energies for graphene lattice vacancy diffusion and merging were reported to be in the range of 1.1-1.3 eV[53] and 1.2-2.1 eV[53] respectively. Table 1 of Yang et al. gives a summary of various methods for computing vacancy formation.[54]

Here, we used two methods to calculate the pair-wise potential directly from the experimental video. One method was based on structure and the other based on velocities. The first method used pairwise distances from video that were used with kernel density estimation (KDE). Gaussian kernels with a bandwidth of 0.2 were used. KDE gives a probability of pairwise distance, P(r). The radial distribution function (RDF), $g(r)$ was subsequently calculated according to the equation: $C_1 \frac{P(r)}{r^2}$, where $C_1$ is a constant determined to ensure $g(r)$ is unity at large $r$. $g(r)$ was then used to calculate the two particle PMF, $w^{(2)}$, using the equation:

$$w^{(2)} = -kT \ln g(r) - C_2 \qquad (1)$$

This approach of deriving the PMF from the RDF is valid only when we have statistical sampling from an equilibrium ensemble.[55] Since the video from the STEM experiment shows ablation and gradual decrease in the particle number, the canonical restriction is not necessarily satisfied here. Nonetheless, we argue that the character of the process suggests significant stationary component and hence this approach establishes a useful baseline for the two-body potential.

The second approach is to assume Langevin dynamics and fit the potential directly. This requires no assumptions about equilibrium. We assumed a model of pairwise potential using a basis set ($\boldsymbol{B}(r)$) of 32 Gaussian functions along with a repulsive term, $U(r)$. The $U(r)$ is given by:

$$U(r) = u\,(r - r_0)(r - r_0)^{12} \qquad (2)$$

where $u(r)$ is a unit step function. The basis set is of the form:

$$\boldsymbol{B}(r) = h_i\, G(s_i, w_i) \cdot \frac{1}{1+r} \qquad (3)$$

where $h_i$ is the scaled height, $s_i$ is the mean and $w_i$ is the standard deviation of the $i^{th}$ Gaussian basis set function. We use overdamped Langevin dynamics to calculate the forces directly from the coordinates. The generalized Langevin equation[56] is given by:

$$m\dot{v} = -\frac{\partial U}{\partial r} + \xi(t) - \gamma v \qquad (4)$$

where $\gamma v$ is the dampening force and $\xi(t)$ is the random noise term. $U$ is the potential energy function, $r$ is the distance and $v$ is the velocity. $\gamma$ is the friction coefficient and $\xi(t)$ is a Gaussian with zero mean. The velocity correlation decay time is given by $\tau = m/\gamma$. For times, $t \gg \tau$, the inertia term, $m\dot{v}$, can be neglected. This case is referred to as the overdamped Langevin.[57] The corresponding equation is given by:

$$v = -\gamma^{-1}\frac{\partial U}{\partial r} + \gamma^{-1}\xi(t) \qquad (5)$$

We have chosen the overdamped Langevin model since the time lapse between two consecutive samples from the experimental video is 17 s, a significant time period compared to the scale of atomic motion. Figure S1 in the supporting information (SI) gives a quantitative evidence of the velocity autocorrelation function calculated from the experimental video. It shows rapid decay of the autocorrelation between $\tau = 0$ and $\tau = 1$ supporting the assumption that the correlation time is smaller than the time difference between consecutive frames. This shows that, like the overdamped Langevin model, there is no momentum correlation between frames. Besides the forces arising from pair-wise interactions, all the contributions from other dynamical processes are incorporated in the random force term in equation (5). Using initial parameter values for $\boldsymbol{B}(r)$, equation (5) and a velocity-Verlet integrator with unit time-step, we generate the predicted coordinates of the atoms in the next timestep. The squared difference between the predicted coordinates $(y')$ and the true coordinates $(y)$ from the next frame of the video is the objective function to be minimized. Since the number of particles in each frame is not constant in the experimental video, we used an index reassignment strategy to trace each atom along the trajectory. To account for the image boundary effects and variable number of particles across the trajectory, we assigned a weight $(\theta_i)$, which is set to either 1 or 0, to each atom. The weight determines if an atom is considered during training. To ensure that all the frames are equally weighted during parameter training the cumulative weight $(\theta)$ for each frame was scaled to a uniform value.

The loss function which is minimized during training is given by:

$$\chi = (y - y')^2\theta + \frac{1}{n_B}\sum_i^N h_i^2 - r_0 \qquad (6)$$

where $n_B$ is the number of Gaussian functions in the basis set (32). This optimizes the potential function parameters. A similar method of learning pair-wise potentials using Gaussian basis set functions has been previously reported.[58] This is also like the use of a spline basis-set seen in force matching.[59]

We validate our method by applying the discussed pair-potential regression strategy using a toy model with two types of particles, A and B, following overdamped Langevin dynamics of known parameters. The masses of A and B were set to those of C and Si, respectively. Lennard-Jones (LJ) potential was used as the potential in the Langevin equation for the toy-system. To generate the synthetic reference trajectory for the toy system, serving as a proxy to for the experimental video, we used the first frame of the experimental video and initialized the positions of the particles for the first frame of the synthetic trajectory. We also tested multi-step integration between each step.[60] These results can be found in the SI.

All the calculations for both the methods, PMF calculation using RDF and pair-potential inference using overdamped Langevin dynamics, were based on the experimental video. The distance values had to be scaled to proper units. We used the first peak in the radial distribution function, corresponding to the C-C graphene bond, for scaling the distance to match the standard C-C graphene bond length of 0.142 nm. Figure 2 shows the RDFs for C-C and C-Si in graphene and the corresponding two particle PMFs, $w^{(2)}{}_{C-C}$ and $w^{(2)}{}_{Si-C}$, calculated using this approach.

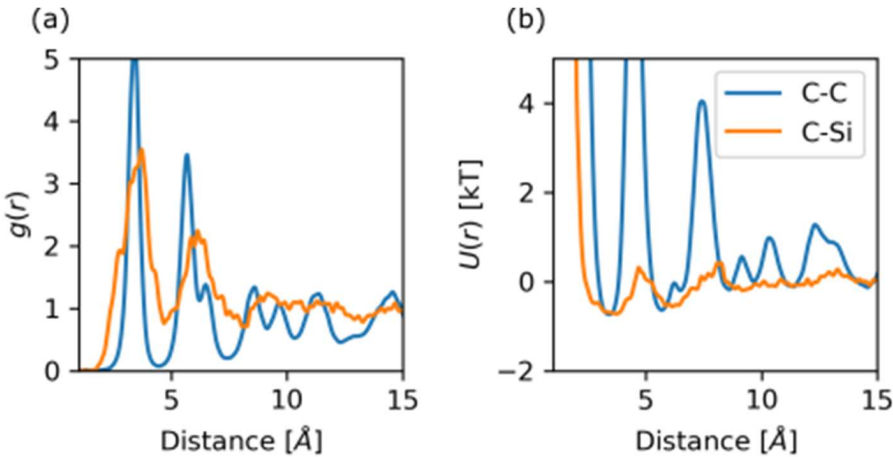

**Figure 2 RDFs and the corresponding PMFs for pair-wise interactions in graphene.** (a) C-C and C-Si RDFs in graphene generated from the processed experimental video. The RDFs were calculated using histograms of C-C and C-Si pair-wise distances and fitting Gaussian KDE

functions to the histogrammed distances. (b) two particle PMFs, $w^{(2)}_{C-C}$ and $w^{(2)}_{Si-C}$, calculated from the RDFs obtained.

Figure 3 compares the two particle PMFs, $w^{(2)}{}_{C-C}$ and $w^{(2)}{}_{Si-C}$, and Si-C and C-C pair-potentials trained the Langevin assumption. The locations of the first repulsive wells, indicated by the vertical lines in Figure 3, are comparable. The later minimums in the potential could be from lack of three-body terms. The Langevin fit is likely better, because it is based on a regression to the observed particle motions. The PMF, as mentioned above, should only agree if ergodic, equilibrium, NVT sampling is observed. These conditions do not hold, especially since the majority of statistics contributing are from the static graphene lattice. To obtain the well-depth in Figure 3, we calibrated the energies according to the C-C Lennard-Jones parameters from Inui and Iwasaki (4.55 meV). This results in an C-Si minimum of 1.5 meV, showing the Si fits less favorably into the graphene lattice than carbon which is expected. Regarding the shape of the potentials, the Langevin method yields less extreme potentials but is still showing large peaks, likely because the graphene lattice structure is being projected onto this two-body potential.

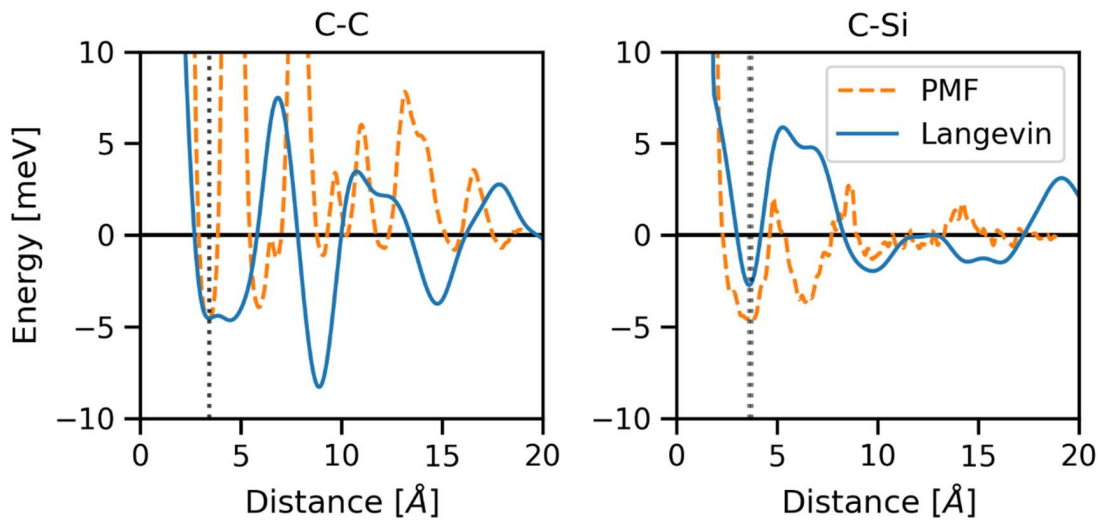

**Figure 3 C-C and Si-C PMFs and pair-wise potentials as obtained by the two different methods.** The result plotted in orange is the pair-wise PMFs obtained using RDF and KDE. The pair-wise potentials in blue were learned using Gaussian basis set functions.

**Conclusions**

In this work we have shown that the videos generated by dynamic STEM experiments can directly be used to reconstruct pair-wise potentials. We have explored two techniques for reconstruction, one based only on structures observed and one based on the motion of the atoms. The use of motion (Langevin) requires less assumptions and give more reasonable approximate two-body pairwise potentials. The structure-based approach (PMF) requires many assumptions and gives unreasonably large forces, mostly because a PMF analysis is inapplicable to the large fraction of static atoms. This work highlights the potential of using state-of-the art characterization techniques, dynamic STEM, in conjunction with computational modeling methods to learn the underlying physics of different phenomena. This warrants further research along the line of the present work.

**Supplementary Material**

Supplementary figures 1-5.

**Data Availability**

The data that support the findings of this study are available from the corresponding author upon reasonable request.

**Acknowledgements**

This effort (electron microscopy, feature extraction) is based upon work supported by the U.S. Department of Energy (DOE), Office of Science, Basic Energy Sciences (BES), Materials Sciences and Engineering Division (O.D., S.J., S.V.K) and was performed and partially supported (MZ) at the Center for Nanophase Materials Sciences, which is a DOE Office of Science User Facility. Theoretical analysis (MC, ADW) is based upon work supported by the National Science Foundation under grants 1764415 and 1751471. The authors are grateful to Dr. R. Unocic for careful reading and commenting upon the manuscript.